\documentclass{elsart}
\usepackage{graphicx}
\usepackage{amssymb}
\usepackage{xspace}
\usepackage{amsmath}
\usepackage{url}
\usepackage{rotating}
\usepackage{cite}

\journal{Astroparticle Physics}

\renewcommand{\vec}[1]{\boldsymbol{#1}}
\newcommand{\unit}[2]{\ensuremath{#1\:\mathrm{#2}}}

\newcommand{\micro}{\mu}

\newcommand{\capt}[2]{\caption{#2}\label{#1}}

\newcommand{\fg}[1]{Fig.\thinspace\ref{#1}}
\newcommand{\eq}[1]{Eq.\thinspace(\ref{#1})}
\newcommand{\tb}[1]{Table\thinspace\ref{#1}}

\newcommand{\secref}[1]{Section \ref{#1}}
\newcommand{\appref}[1]{Appendix \ref{#1}}

\newcommand{\standardgraphic}[3][width=\columnwidth]{%
  \begin{figure}[ptb]%
  \centering%
  \includegraphics[#1]{#2.eps}%
  \capt{#2}{#3}%
  \end{figure}%
}

\newcommand{\multicolumngraphic}[3][width=\textwidth]{%
  \begin{figure*}[ptb]%
  \centering%
  \includegraphics[#1]{#2.eps}%
  \capt{#2}{#3}%
  \end{figure*}%
}

\newcommand{\eg}{\textit{e.g.}\ }

\begin{document}
\begin{frontmatter}

\title{A phenomenological model of the muon density profile on the ground of very inclined air showers}
\author[RWTH]{H. P. Dembinski\corauthref{cor}},
\ead{dembinski@physik.rwth-aachen.de}
\author[LPNHE]{P. Billoir},
\author[IPN]{O. Deligny}, and
\author[RWTH]{T. Hebbeker}
\corauth[cor]{Corresponding author.}

\address[RWTH]{III. Physikalisches Institut A, RWTH Aachen University, Germany}
\address[LPNHE]{LPNHE (Universit\'{e}s Paris 6 \& 7, CNRS-IN2P3), Paris, France}
\address[IPN]{IPN, CNRS/IN2P3-Universit\'{e} Paris Sud, Orsay, France}

\begin{abstract}
Ultra-high energy cosmic rays generate extensive air showers in Earth's atmosphere. A standard approach to reconstruct the energy of an ultra-high energy cosmic rays is to sample the lateral profile of the particle density on the ground of the air shower with an array of surface detectors.

For cosmic rays with large inclinations, this reconstruction is based on a model of the lateral profile of the muon density observed on the ground, which is fitted to the observed muon densities in individual surface detectors. The best models for this task are derived from detailed Monte-Carlo simulations of the air shower development. We present a phenomenological parametrization scheme which allows to derive a model of the average lateral profile of the muon density directly from a fit to a set of individual Monte-Carlo simulated air showers. The model reproduces the detailed simulations with a high precision. As an example, we generate a muon density model which is valid in the energy range $\unit{10^{18}}{eV} < E < \unit{10^{20}}{eV}$ and the zenith angle range $60^\circ < \theta < 90^\circ$.

We will further demonstrate a way to speed up the simulation of such muon profiles by three orders of magnitude, if only the muons in the shower are of interest.
\end{abstract}


\end{frontmatter}

%
%
%
\section{Introduction}
\textbf{U}ltra-\textbf{h}igh \textbf{e}nergy \textbf{c}osmic \textbf{r}ays (UHECRs) are cosmic rays with energies above 1 EeV $=$ \unit{10^{18}}{eV}. They have been under study for several decades, still their origin is not well known. Information about the origin is encoded in the energy spectrum~\cite{Schussler2009}, the mass composition~\cite{Unger2007} and a possible anisotropy of their arrival directions~\cite{Abraham2007}, which can be measured experimentally.

The huge energy and the low flux (roughly 1 particle per km$^2$ per year above 10 EeV) make a direct measurement of the momentum and energy of a UHECR with balloon or satellite experiments unfeasible. Instead, they are observed indirectly with ground-based detectors that use Earth's atmosphere as a large calorimeter. The interactions of the UHECR with atmospheric nuclei generate an extensive particle shower which is sampled by these detectors.

One realization of such a detector is a large surface array of particle counters. The array samples the lateral profile of the particle density of the shower which consist mainly of photons, electrons, and muons. The arrival direction of the UHECR can be obtained rather directly from the measured arrival time of the shower front in individual particle counters. The reconstruction of the energy $E$ of the cosmic ray is more complex and requires to fit a model of the lateral particle distribution around the shower axis to the measured particle counts.

Most surface array experiments concentrate on showers with inclinations less than $60^\circ$. For such showers, the particle distribution is radially symmetric in good approximation and well described by comparably simple empirical models of the NKG-type~\cite{Kamata1958,Greisen1960}. At larger inclinations, the effect of the geomagnetic field on the particle distribution cannot be neglected. The symmetry becomes broken and the NKG-type models fail to describe the particle distribution. 

In an ideal surface array, these very inclined showers constitute \unit{25}{\%} of the number of arriving events. Recovering them yields a significant gain in the event statistics of the experiment. The ability to reconstruct very inclined showers also increases the field of view of the detector and thus the total observable region of the sky. This is particularly relevant for anisotropy searches.

It was first demonstrated with the Water-Cherenkov detectors of the Haverah Park experiment~\cite{Edge1973} that the energy $E$ of cosmic rays at large zenith angles $\theta > 60^\circ$ can be derived from the total number of muons $N_\mu$ which arrive at the ground~\cite{Ave2000a,Ave2000b}. The same approach is now used by the Pierre Auger Observatory~\cite{Vazquez2009}. The number of muons $N_\mu$ on the ground is obtained from a fit of a model of the average lateral profile of the muon density\footnote{The term ``muon density'' in this article refers to the time-integrated particle flux through the ground plane initiated by an air shower.} $n_\mu$ on the ground to the measured signals. This fit exploits the following factorization
\begin{equation}
  n_\mu \simeq N_\mu(E,A,\theta) \times f_\mu(x,y,\theta,\phi),
\end{equation}
whereas $N_\mu$ is the number of muons on the ground which depends only on the energy $E$, mass $A$, and inclination $\theta$ of the cosmic ray, while $f_\mu$ is a normalized lateral profile of the muon density which depends only on the ground coordinates $(x,y)$ and the shower direction $(\theta,\phi)$. The normalized profile $f_\mu$ also depends on the properties of the observation site like the ground altitude, the geomagnetic field and the atmosphere, but those are considered fixed here. In simple terms, the factorization says that the \emph{shape} of the lateral profile remains the same for all showers arriving from a certain direction in very good approximation, while its \emph{amplitude} carries all the information about the energy $E$ and mass $A$ of the cosmic ray. This invariance of the profile shape is called \emph{shower universality}.

The universality is very useful, because $f_\mu$ is too complex to be fitted from the data sampled by the surface array on an event-by-event basis. It has to be modeled. However, if $f_\mu$ is predicted by a model, the reconstruction of $N_\mu$ reduces to a fit of three parameters: the two intersection coordinates of the shower axis with the ground and the amplitude $N_\mu$.


This reconstruction approach even works if the particle counters of the surface detector cannot distinguish between different species of charged particles. Very inclined air showers arrive in a very late stage of their development on the ground, where the only remaining electromagnetic particles in the shower are generated by the muons themselves, mostly via decay. Therefore, signals generated by such old showers remain proportional to the local muon density $n_\mu$, because the electromagnetic particles only enhance the signal by a constant factor in first approximation~\cite{Valino2010}.

Models of the normalized profile $f_\mu$ are currently derived from detailed Monte-Carlo simulations of extensive air showers, which represent best our current theoretical knowledge. However, it is not feasible to make a Monte-Carlo simulation of $f_\mu$ for every possible shower direction $(\theta,\phi)$, simply because these simulations consume considerable computing time and and storage space. The established solution to this issue is a semi-analytical model of $f_\mu$, which is based on detailed simulations but predicts the azimuthal dependency of $f_\mu$ analytically~\cite{Ave2000a}. The semi-analytical model is instructive and reproduces the detailed simulations well, but still deviates somewhat from full air shower simulations since some effects are neglected in the analytical part of the model.

In this article, we follow a different path and propose a phenomenological parametrization of $f_\mu$ which can be fitted directly to a set of simulated air showers. The parametrization is not derived from an analytical theory, it only exploits some general principles of the air shower development. The approach is valid up to at least \unit{4}{km} from the shower axis in the energy range $\unit{10^{18}}{eV} < E < \unit{10^{20}}{eV}$ and the zenith angle range $60^\circ < \theta < 88^\circ$ and reproduces the output of detailed simulations better than the semi-analytical model. As a consequence, our model should lead to smaller biases in the reconstructed muon number $N_\mu$ on the ground.

The parametrization procedure does not require a specific distribution of simulated showers over the parameter range $(\theta,\phi)$ of interest as input as long as that range is sufficiently covered with simulated showers. It can therefore be applied to many existing air shower libraries. As a consequence of the shower universality, the phenomenological parametrization of $n_\mu$ is practically independent of the nature of the primary particle, except for a global factor.

In this article, we derive and motivate the parametrization. It is then applied to a set of 1800 simulated air showers as an example and in order to demonstrate its precision. On a related side note, we point out an efficient way to speed up the detailed simulation of $n_\mu$-profiles, which can then be used as input of the parametrization. In particular, we show that exploiting the azimuthal symmetry of the muon profile at the production point and neglecting the calculation of the electromagnetic cascade leads to computation times much smaller than conventional Monte-Carlo simulations.

The article is organized as follows. In \secref{sec_inclined_shower_discussion}, we review some general features of very inclined air showers in order to motivate the parametrization approach. The library of air showers used for the example application and the fast simulation approach are described in \secref{sec_mc_prod}. The parametrization is then discussed in \secref{sec_map_generation} and its precision demonstrated. A summary of the results is given in \secref{sec_conclusion}.

\section{Very inclined air showers: the general picture}\label{sec_inclined_shower_discussion}

The physics (see \eg ref.~\cite{Gaisser1990,Ave2000a,Ave2000b,Billoir2001,Ave2002,Cazon2004,Matthews2005,Dembinski2009}) which influence the muon component of very inclined air showers are reviewed in this section.

The lateral profile of the muon density $n_\mu$ on the ground has different properties for standard showers ($0^\circ < \theta < 60^\circ$) and for very inclined showers ($60^\circ < \theta < 90^\circ$). Standard showers are dominated by photons and electrons generated in the hadronic interactions. The flux of particles through the shower front plane is almost radially symmetric with respect to the shower axis.

Very inclined showers travel longer through the atmosphere. As the shower arrives at the ground, the electromagnetic component generated by hadronic interactions is already absorbed. What remains is a muon shower in first approximation. The lateral profile of the muon density gets deformed by the geomagnetic field, atmospheric attenuation, and geometrical effects, so that the particle flux through the shower front plane is no longer radially symmetric. These asymmetries become very large as $\theta$ approaches $90^\circ$.

The primary electromagnetic component coming from the decay of neutral pions is already almost absorbed when the shower reaches the ground at $\theta \approx 60^\circ$ and can be considered extinct at $\theta \gtrsim 70^\circ$ (see also \fg{shower_example1}). A non-negligible electromagnetic component is still detectable on the ground at all angles which is produced by the muons themselves. The muons produce photons and electrons mostly through decay, but also via bremsstrahlung, $e^{+} e^{-}$-production and delta rays along their path through the atmosphere. In some sense, they are surrounded by electromagnetic sub-showers with a moderate lateral extension -- typically a few tens of meters.

Therefore, as long as the muon density $n_\mu$ is larger than about \unit{0.1}{m^{-2}}, these electromagnetic sub-showers overlap result in a continuous halo. The electromagnetic particles arrive together with the muons within a few tens of nanoseconds and do not increase the longitudinal thickness of the shower front. However, they have to be considered in particle detectors that cannot distinguish between muons and the electromagnetic halo. We will ignore the impact of electromagnetic particles in the following. We assume that either the detectors distinguish between different particle species, so that the electromagnetic component can be ignored, or that the close proportionality between electromagnetic and muonic components in very inclined showers is exploited in the data analysis in order to estimate the muon density from the mixed signal~\cite{Ave2000b,Valino2010}.

\subsection{Conventions}

Quantitative calculations and detailed simulations of extensive air showers depend on the observation site. Important are the local geomagnetic field $\vec{B}$, the altitude of the ground $h_g$ above sea level and the air density profile of the air above the site. In this article, we do all calculations for the site of the Southern Pierre Auger Observatory in Malarg\"{u}e, Argentina.

The ground plane altitude is taken as \unit{1425}{m}. The geomagnetic field $\vec{B}$ is treated as a constant field\footnote{The geomagnetic field currently varies by about $1^\circ$ in direction and \unit{2}{\%} in magnitude over \unit{10}{years} in Malarg\"{u}e~\cite{ngdc_url}.}:
\begin{equation*}
  B = \unit{24.6}{\micro T}, \quad \delta_B = 4.2^\circ, \quad \theta_B = -35.2^\circ,
\end{equation*}
whereas $\delta_B$ and $\theta_B$ are the geomagnetic declination and inclination.

An average profile of the air density over the site is approximated by the US standard atmosphere~\cite{NASA1976}.

The muon profiles derived in this article are supposed to be comparable with experimental measurements and therefore need to regard the energy threshold of the applied particle counters. Scintillators detect practically all muons; their threshold for Cherenkov emission in water is about 50 MeV. However the threshold for buried or shielded detectors can reach a few GeV. The simulations shown in this article are done for a muon energy threshold
\begin{equation*}
  E_\text{thres}^\mu = 0.145 \text{ GeV}.
\end{equation*}
The results of this article remain qualitatively correct if it is increased up to a few GeV, but quantitatively they depend on the threshold. The dependency of muon profiles on the observation site and the muon energy threshold is not mentioned explicitly in the rest of the article.

\standardgraphic[width=0.7\textwidth]{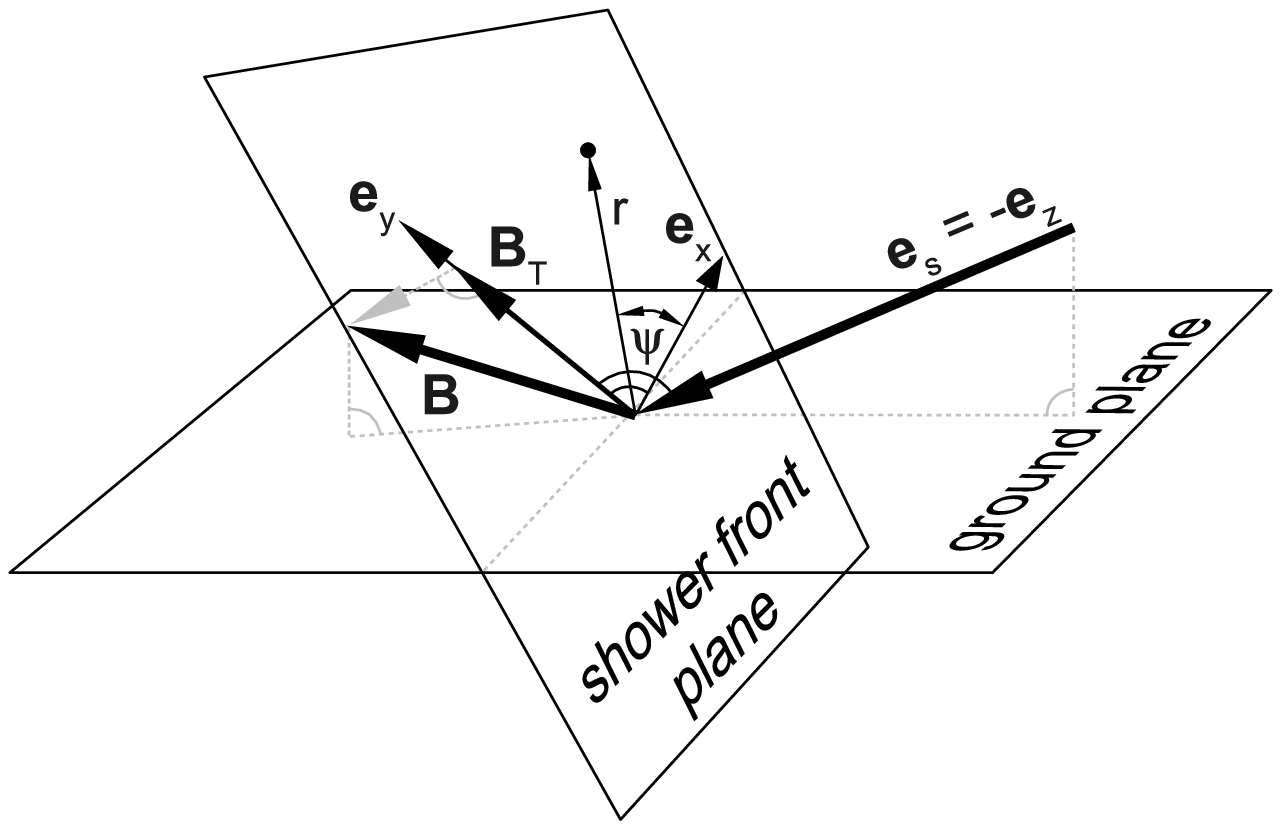}{The shower front plane coordinate system~\cite{Ave2000a}: $\vec{e}_z$ is anti-parallel to the momentum vector of the shower, $\vec{e}_y$ is parallel to $\vec{B}_T$, the component of the geomagnetic field projected into the shower plane. $r$ and $\psi$ are polar coordinates in the shower front plane.}

In order to discuss the lateral profile of an extensive air shower, it is useful to introduce
a special coordinate system: the shower front plane coordinate system (see
\fg{coordinate_system}). The shower front plane is perpendicular to the shower
axis. All observations are still done in the ground plane, but the coordinates are projected onto this
plane. The projection restores some of the principal symmetries of the shower profile. We will refer to this coordinate system during the rest of the article.

Finally, we use the convention of the Pierre Auger Observatory for the azimuthal angle $\phi$, where a shower arriving from the geographic East equals to $\phi=0^\circ$, and a shower from the geographic North equals to $\phi=90^\circ$.

\subsection{Development of the muon component}

\multicolumngraphic{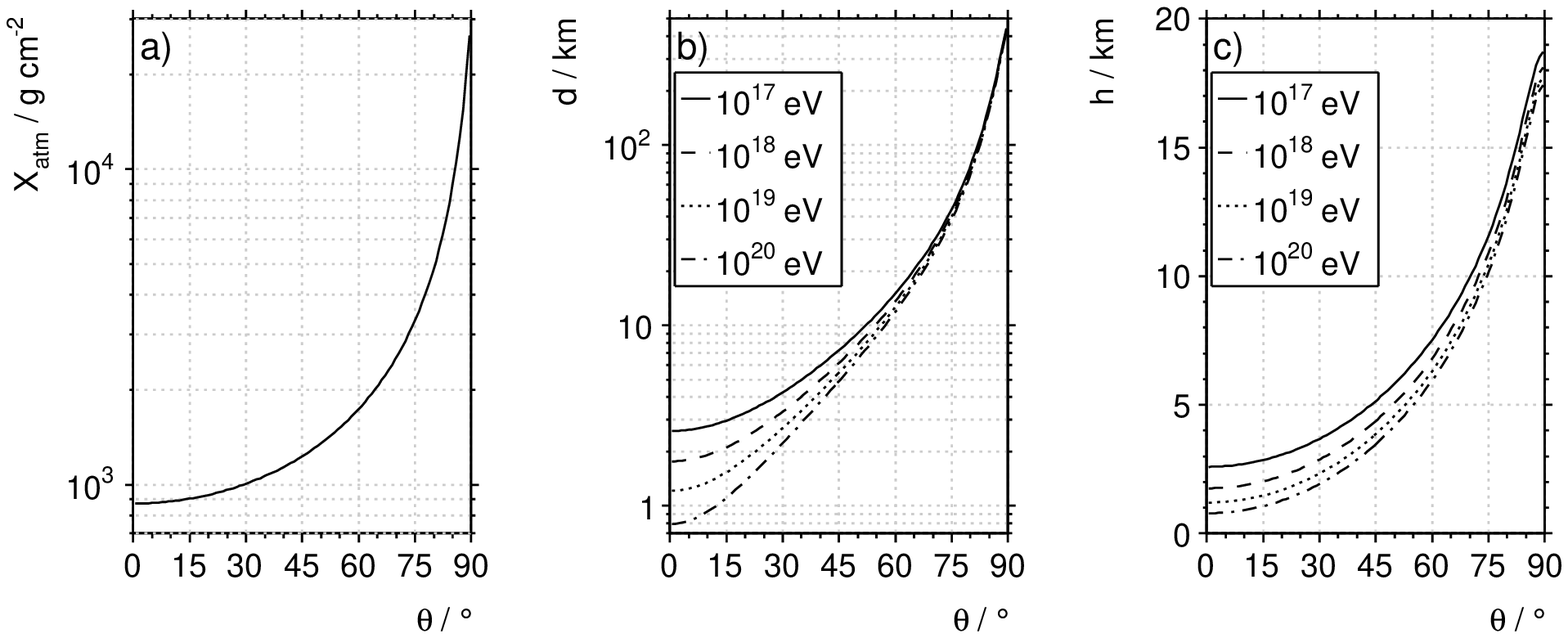}{As a function of the zenith angle $\theta$ at the ground altitude of 1425 m, we show: a) the integrated atmospheric slant depth $X_\text{atm}$ along the shower path, b) the distance $d$ between the shower maximum and the impact of the core on the ground, and c) the altitude $h$ of the electromagnetic shower maximum above the ground level. The latter are shown for different energies, because the depth $X_\text{max}$ of the shower maximum depends on the energy $E$. For the calculations of the slant depth $X$ along a shower path in a curved atmosphere see \eg \cite{Gaisser1990}. A parametrization of $X_\text{max}(E)$ is taken from \cite{Unger2007}.}

An air shower induced by a proton or a nucleus first produces a hadronic cascade which mostly produces charged and neutral pions in each step. The neutral pions decay almost immediately into two photons and feed an electromagnetic component. The decay of the charged pions feeds a muon component. The total atmospheric slant depth $X_\text{atm}$ increases with the zenith angle $\theta$; for example, for an observer at 1400 m above sea level, when $\theta$ rises from $0^\circ$ to $90^\circ$, the depth down to the ground increases from \unit{870}{g\, cm^{-2}} to \unit{31000}{g\, cm^{-2}}  (see \fg{depth_d_h}).

\begin{sidewaysfigure}
  \centering
  \includegraphics[width=\textwidth,height=11cm,keepaspectratio]{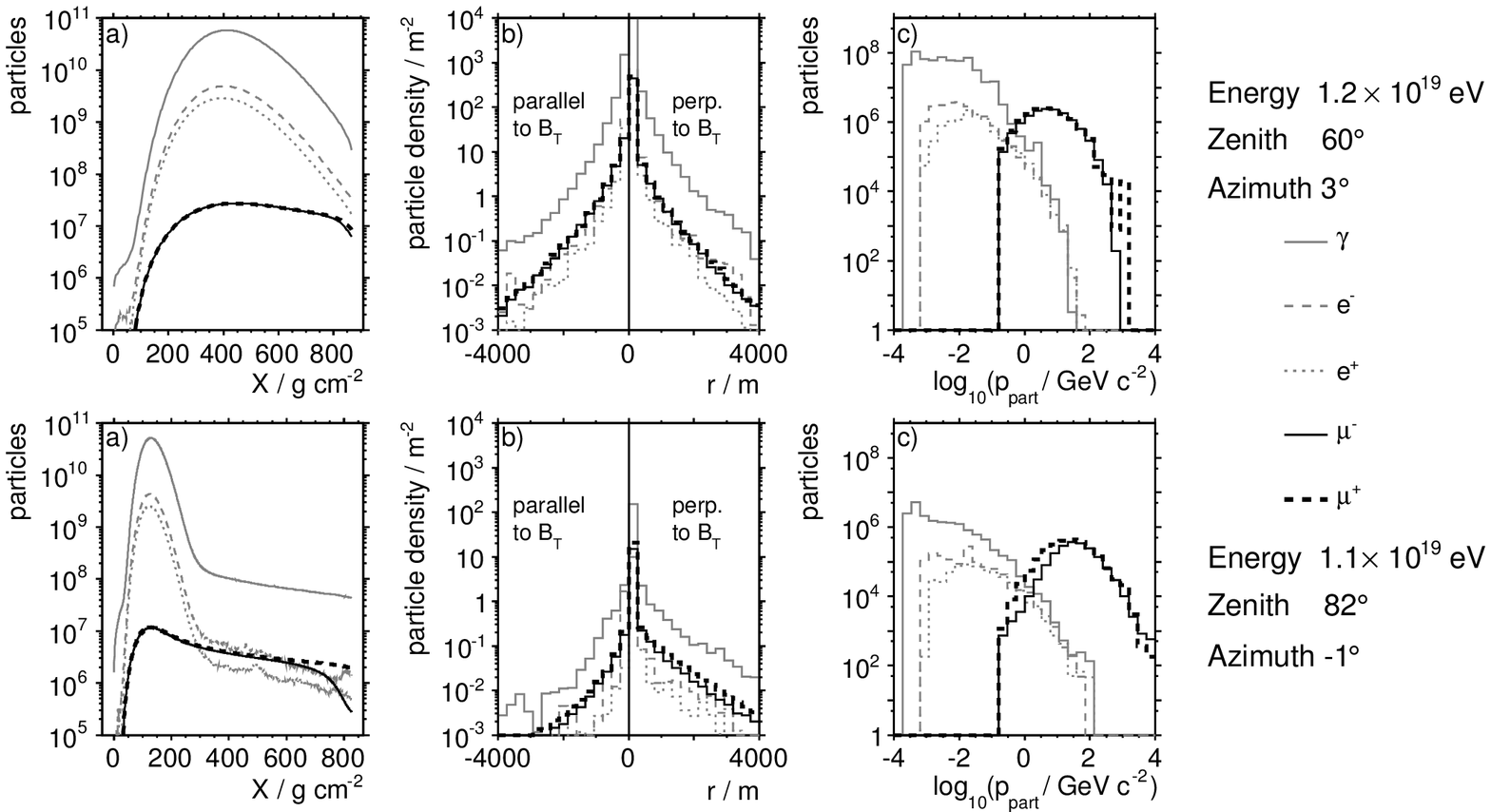}
  \capt{shower_example1}{Comparison of two inclined proton showers, simulated
with CORSIKA using QGSJet-II and FLUKA for the hadronic interactions~\cite{corsika,qgsjet2,fluka}. The energies, zenith and azimuth angles of two showers are given in the legend right of the plots. The plots show a) the total particle number vs. vertical atmospheric depth, b) the particle number per tank on the ground vs. radial distance from the shower axis, and c) the momentum spectrum on the ground. In b), $r>0$ relates to particles collected within $\pm 10^\circ$ around the direction perpendicular to the projected geomagnetic field $\vec{B}_T$ and $r<0$ (to be taken in absolute value) relates to particles collected within $\pm 10^\circ$ around the direction parallel to $\vec{B}_T$. In all plots, only particles above certain energy thresholds are shown: $E_\text{thres} = \unit{250}{keV}$ for electromagnetic particles and $E_\text{thres} = \unit{0.1}{GeV}$ for everything
else. The low energy photon peak in c) is caused by $e^+/e^-$ annihilation. The particle depletion along $\vec{B}_T$ in b) for $\theta = 82^\circ$ is the result of geomagnetic deflection. The rapid decay of the muon number at large slant depths is artificial and caused by the lateral extension of the muon profile and the fact that CORSIKA removes particles from the longitudinal profile which hit the observation level.}
\end{sidewaysfigure}

At an inclination $\theta > 60^\circ$, the total atmospheric depth $X_\text{atm}$ is more than twice the depth $X_\text{max}$ of the air shower maximum, which ranges between \unit{650}{g\,cm^{-2}} and \unit{800}{g\, cm^{-2}} between \unit{10^{18}}{eV} and \unit{10^{20}}{eV}~\cite{Unger2007}, depending on the energy $E$ and mass $A$ of the cosmic ray. $X_\text{atm}$ is more than three times the depth of the end of the hadronic cascade, where most muons are produced (see \fg{shower_example1}). As a consequence, the electromagnetic cascade is fully extinguished at ground level. Only the most energetic muons survive, accompanied by an electromagnetic halo produced by the decay and other radiative processes of the muons.

The air density for a given slant depth decreases with increasing $\theta$. The pions decay when their decay length becomes comparable to their interaction length. The latter is inversely proportional to the air density and thus pions at the end of the hadronic cascade tend to decay at higher energy at larger $\theta$. Muons inherit \unit{80}{\%} of the energy of their parents on average. Their production energy therefore increases from \unit{20}{GeV} to \unit{100}{GeV} as $\theta$ increases from $60^\circ$ to $90^\circ$.

The angular spread of the muons up to this point is mainly caused by the transverse momentum $p_T$ inherited from the parental pions and the decay angle between the pion and the muon~\cite{Ave2000a}. An additional small random deflection is a caused by the kinematics of the decay. Both effects scale as the inverse of the energy. The radial offset of the pions from the shower axis is only of the order of a few \unit{10}{m} and does not contribute significantly to the lateral distribution of the muons observed on the ground at $r \gtrsim \unit{100}{m}$~\cite{Cazon2004}.

After their production, muons are affected by ionization and radiative energy losses, decay, multiple scattering, and geomagnetic deflections. Below 100 GeV, the energy loss is mainly due to ionization, about \unit{2}{MeV\,g^{-1}\,cm^2}~\cite{Amsler2008}, which translates to about \unit{2}{GeV} (\unit{50}{GeV}) for a shower at $\theta = 60^\circ$ ($90^\circ$). The decay length is proportional to the energy, for a \unit{10}{GeV} muon it is \unit{66}{km}. Due to the increase of the production energy with $\theta$, the decay length of the average muon is always larger than the distance $d$ from the production point to the ground (see \fg{depth_d_h}). This explains why a significant fraction of the muons reaches the ground at all zenith angles.

Multiple scattering in the electric field of air nuclei randomizes the directions of muons to some degree and erases small scale correlations in the lateral profile of the muon density $n_\mu$. The effect remains small for the total angular divergence of the muons from the shower axis up to about $\theta \approx 80^\circ$ where multiple scattering becomes the dominant source of the angular divergence apart from geomagnetic deflections.

Geomagnetic deflections $\delta x$ for muons in air showers can be approximated as~\cite{Ave2000a,Billoir2001}
\begin{equation}\label{eq_geomag_defl}
  \delta x \simeq \frac{e\,B_\text{T}\,d^2} {2 E_\mu/c},
\end{equation}
where $e$ is the elementary charge, $d$ the distance between the muon production point and the ground along the shower axis, $E_\mu$ is the muon energy, and $B_\text{T}$ is the perpendicular component of the geomagnetic field with respect to the muon direction. Without the geomagnetic field $\vec{B}$, the air shower development is symmetric in $\phi$. The dependency of the perpendicular component $B_\text{T}(\theta,\phi)$ on $\phi$ breaks this symmetry.

In the extreme case, a \unit{10}{GeV} muon at a zenith angle of $\theta = 60^\circ$ ($80^\circ$) gets a lateral displacement of about \unit{40}{m} (\unit{1700}{m}) after a propagation distance of \unit{10}{km} (\unit{66}{km}) to the ground level. The impact on the shape of the lateral profile is significant. Still, since $\delta x \ll d$, the total number of muons $N_\mu$ on the ground does not depend on the azimuth angle $\phi$.

\standardgraphic[width=0.7\textwidth]{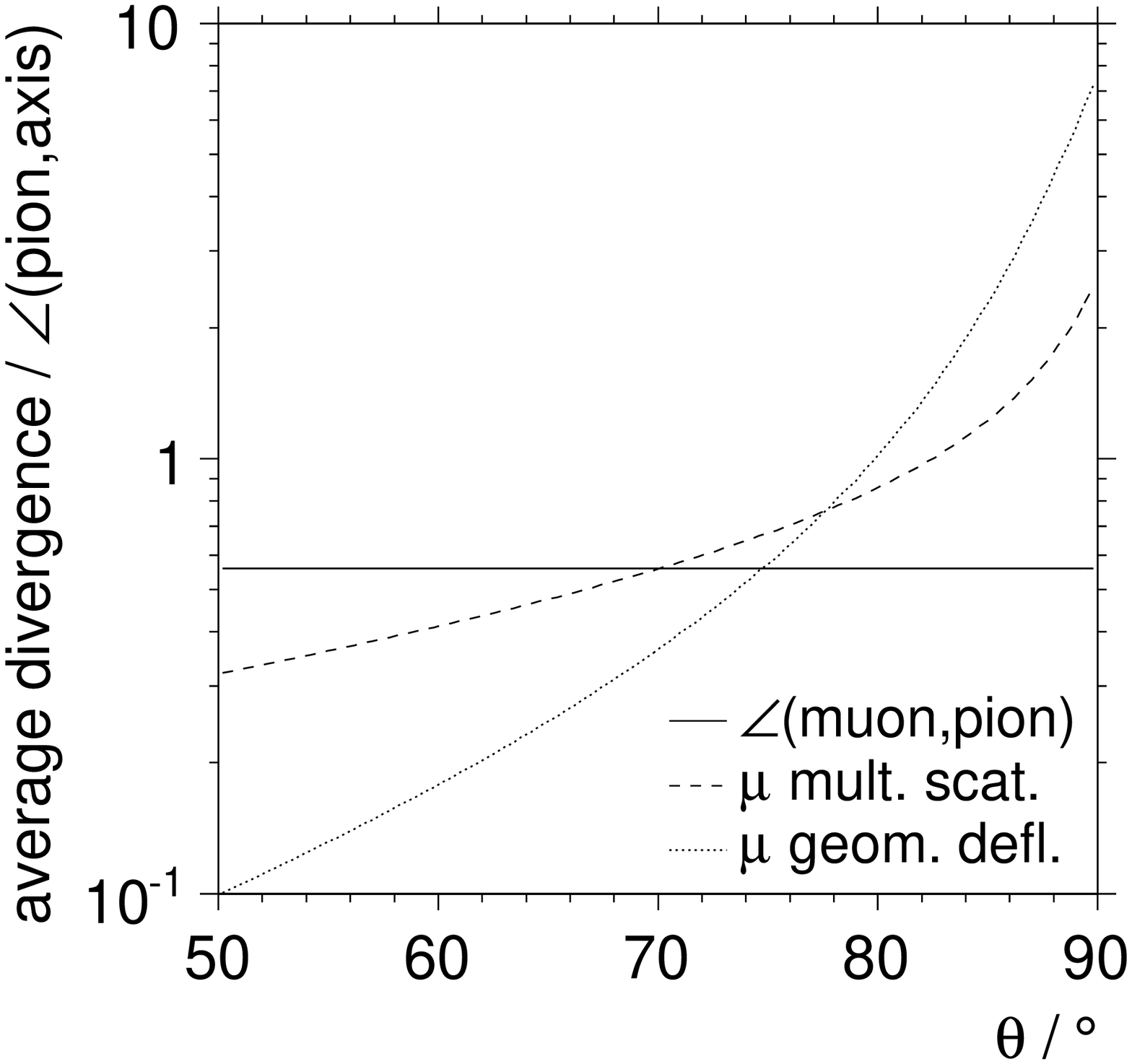}{Average divergence of muons from the shower axis at ground level, normalized to the average angle between the parent pion and the shower axis, as a function of the zenith angle $\theta$ of the shower. The normalized pion-to-muon angle (solid line) is determined by the kinematics of the decay and independent of $\theta$; the normalized angle due to multiple scattering (dashed line) depends on the accumulated slant depth up to ground; the normalized magnetic deflection (dotted line, assuming the maximum effect in Malarg\"{u}e) is proportional to the square of the traveled geometrical distance.}

\fg{divergence_muon} shows the relative magnitude of all effects that contribute to the lateral spread of the muons observed on the ground. The influence of the geomagnetic field becomes important at $\theta \approx 70^\circ$ and dominant at $\theta \gtrsim 80^\circ$.

Muons in the early arriving part of the shower travel shorter distances and have smaller inclinations with respect to the ground plane than muons in the late arriving part. These effects cause an early-late asymmetry which also needs to be taken into account if the lateral profile of the muon density $n_\mu$ on the ground is considered.

\multicolumngraphic{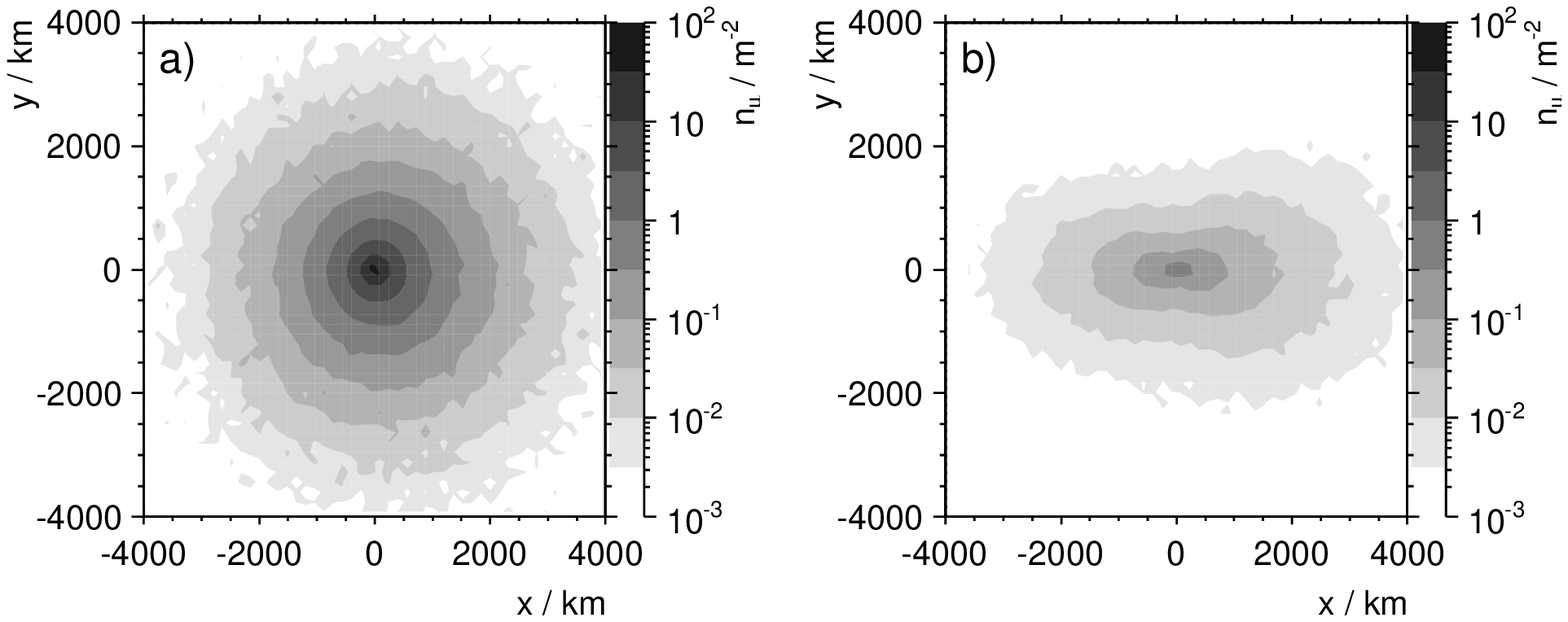}{Lateral profiles of the muon density $n_\mu$ in the shower front coordinate system of the two showers from \fg{shower_example1}, a) $\theta \approx 60^\circ$, b) $\theta \approx 82^\circ$. The azimuthal angles of the showers are such that the geomagnetic deflections are close to the maximum. The radial symmetry around the shower axis is still mostly intact in a), but clearly broken by geomagnetic deflections in b). The left-right asymmetry corresponds to the early-late asymmetry mentioned in the text.}

The loss of radial symmetry in the shower front plane due to the geomagnetic deformation at very large inclinations and the early-late asymmetry are illustrated by \fg{shower_example2}.

\multicolumngraphic{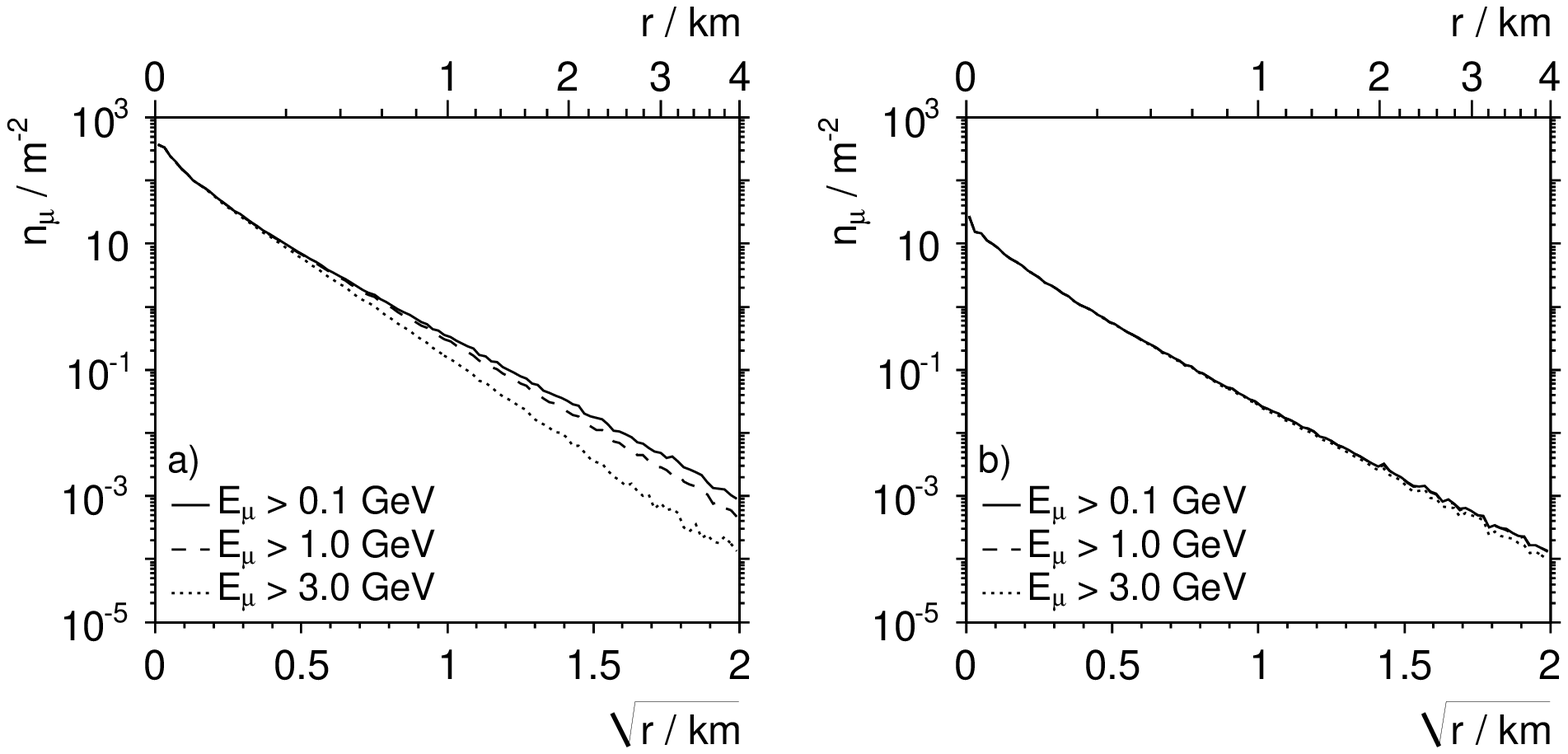}{Same as \fg{shower_example2}, but showing the $n_\mu$ averaged over the polar angle as a function of $r^{1/2}$ in the shower front coordinate system. Different thresholds applied to the muon energy $E_\mu$ show the same approximate functional behaviour $\ln n_\mu \approx \text{const.} + \alpha_1 \, r^{1/2}$.}

Finally, we point out an empirical approach to describe the lateral profile of the muon density $n_\mu$ on the ground in very inclined air showers. We observe that
\begin{equation}\label{eq_nmu_vs_r}
  n_\mu = \frac{\text{d} N_\mu}{\text{d} x \text{d} y} \propto \exp(-\sum_{k=1}^K \alpha_k \, r^{k/2}),
\end{equation}
describes the output of detailed simulations remarkably well. The coefficients $\alpha_k$ depend on the energy threshold $E_\mu^\text{thres}$ of the detectors, but only weakly on the properties of the primary cosmic ray. Even with $K=1$, \eq{eq_nmu_vs_r} is a good first approximation. \fg{dep_thres} demonstrates this.

\eq{eq_nmu_vs_r} has a structure very different from the classical formulas of the NKG-type that are typically used to describe the lateral profile of the muon density~\cite{Greisen1960,Ave2000a,Valino2010}:
\begin{equation}\label{eq_nmu_vs_r_nkg}
  n_\mu \propto r^{\alpha_1'} \, (1 + \alpha_2' \, r)^{\alpha_3'},
\end{equation}
whereas the $\alpha_k'$ are another set of coefficients. Formulas of the NKG-type diverge at $r = 0$.

The choice of a NKG-type formula is generally not motivated by a deeper theory. Kamata and Nishimura theoretically derived a formula of this form for the lateral profile of a purely electromagnetic shower~\cite{Kamata1958}, but the angular divergence in their work is based solely on Coulomb scattering. This is a very good approximation for electromagnetic showers, but not for muonic showers, as shown earlier.

\standardgraphic[width=0.7\textwidth]{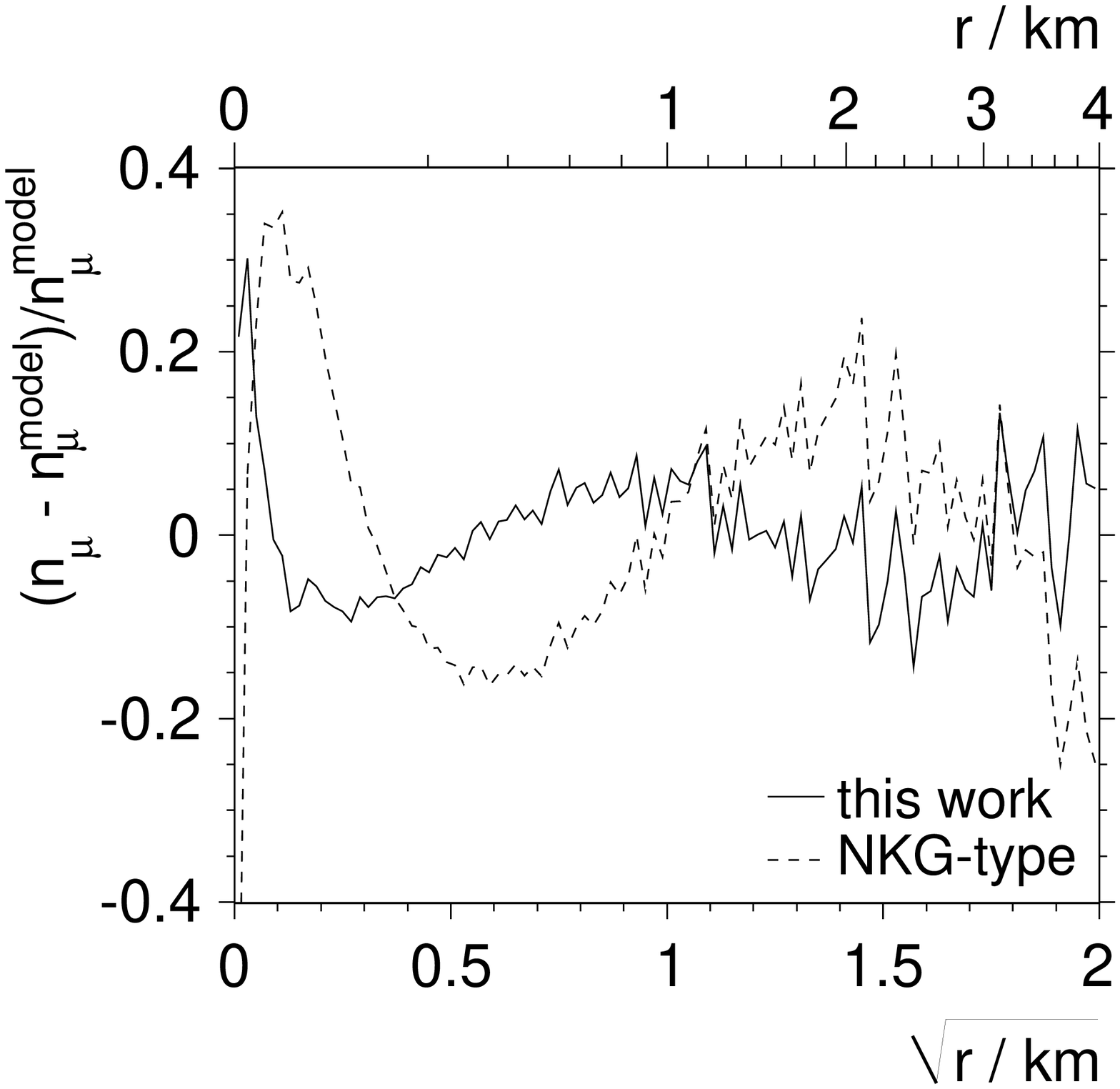}{The lines represent the residuals around fits of \eq{eq_nmu_vs_r} with $K=3$ (this work) and \eq{eq_nmu_vs_r_nkg} (NKG-type) to the lateral profile of the muon density $n_\mu$ from \fg{dep_thres} a) with $E_\mu > \unit{0.1}{GeV}$. \eq{eq_nmu_vs_r} approximates the simulated profile better. Its residuals remain around \unit{5}{\%} up to distances of \unit{10}{m} to the shower axis.}

\eq{eq_nmu_vs_r} always remains finite. It fits better to detailed simulations than \eq{eq_nmu_vs_r_nkg} with the same number of coefficients. An example is shown in \fg{fit_power}. The accuracy of the approximation can be improved by increasing $K$. We note, that \eq{eq_nmu_vs_r} can also be applied well to the lateral profile of the electron or photon density.

The form of \eq{eq_nmu_vs_r} also allows a convenient way of fitting the coefficients $\alpha_k$. We will exploit this and the guaranteed smoothness of the lateral profile due to the multiple scattering in \secref{sec_map_generation}.

\subsection{Energy scaling and shower universality}
\label{sec_energy_scaling}

It was discussed in the previous section that the distributions of the muon energy and the angular divergence from shower axis are a function of the air
density at the altitude $h$ and the slant depth $X_\text{atm} - X_\text{max}$ and distance $d$ from production point to the ground. The amount of geomagnetic deflection also depends on the distance $d$.
 
In very inclined showers with $60^\circ < \theta < 90^\circ$, $X_\text{atm}$ is much larger than $X_\text{max}$. Furthermore, $X_\text{max}$ depends only logarithmically on $E$ and $A$~\cite{Matthews2005,Unger2007}. Thus, $X_\text{atm} - X_\text{max}$, $h$, and $d$ vary only little with the cosmic ray energy in the range $\unit{10^{18}}{eV} < E < \unit{10^{20}}{eV}$ (see \fg{depth_d_h}).

As a consequence, the muon density profile $n_\mu$ factorizes in good approximation into a normalized profile $f_\mu$ and the total number of muons $N_\mu$ on the ground (see also ref.~\cite{Ave2000a,Ave2000b,Ave2002,Dembinski2009}):
\begin{equation}\label{eq_muon_density_factorized}
  n_{\mu}(x,y,\theta,\phi,E,A) \simeq N_{\mu}(\theta,E,A) \times f_{\mu}(x,y,\theta,\phi) \text{ for }\theta>60^\circ.
\end{equation}
The normalized profile $f_\mu$ depends on the momentum distribution of the muons at the production point and the propagation effects to the ground. Both are are approximately independent of the energy $E$ and mass $A$ of the cosmic ray due to the very slow variation of $X_\text{atm} - X_\text{max}$, $h$, and $d$. This invariance property is called shower universality. The total number of muons $N_\mu$ also does not depend on the azimuth angle $\phi$, as stated earlier.

If the profile $n_\mu$ is the one seen by an array of shielded detectors with a significant muon energy threshold, $f_{\mu}$ should be chosen such that it includes the inefficiency of the detector to observe low energy muons. By doing so, $N_{\mu}$ can still be identified with the physical number of muons that arrive on the ground. 

\standardgraphic[width=0.7\textwidth]{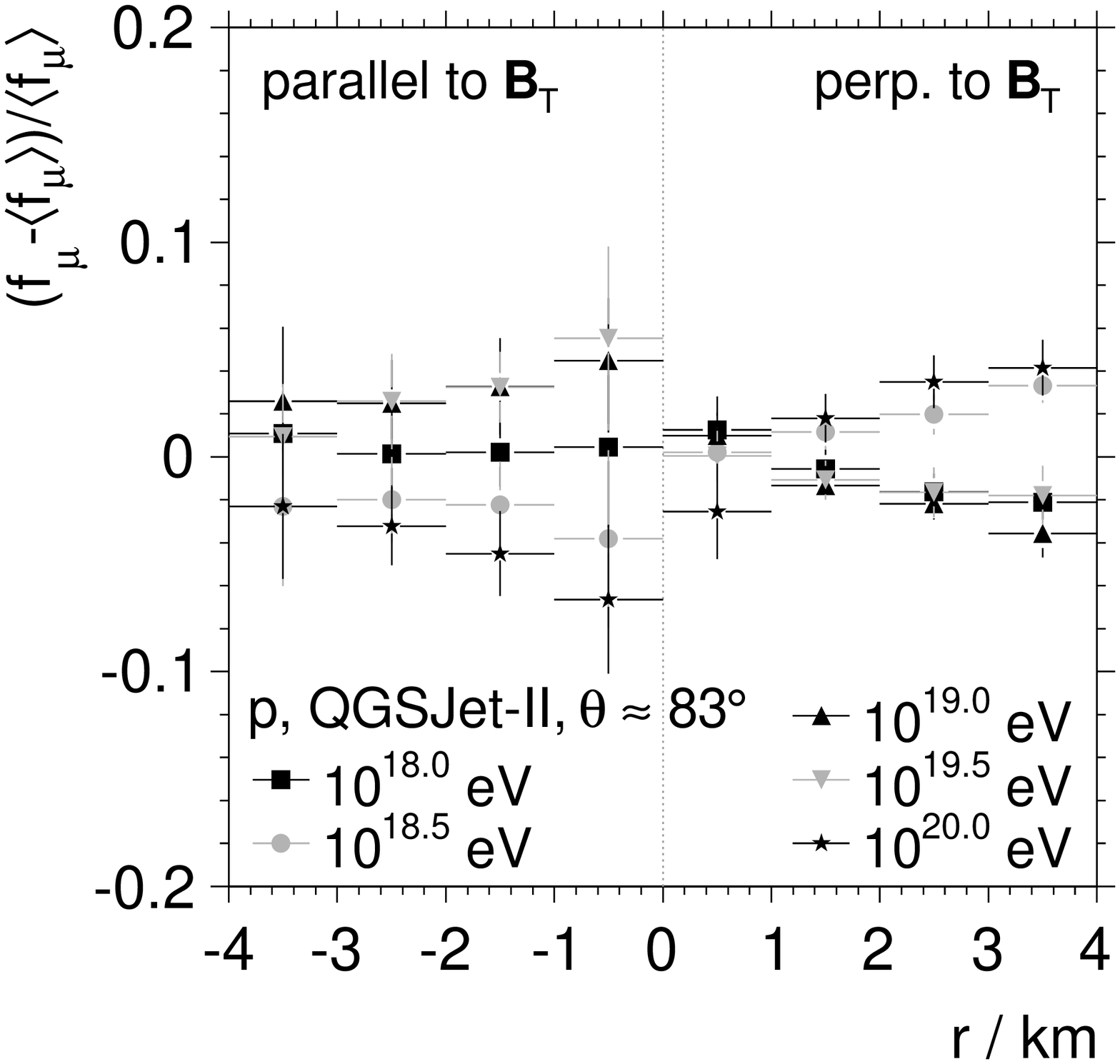}{Variation of the normalized density profile of muons $f_\mu$ at ground level as a function of the cosmic ray energy $E$ for proton showers simulated with the hadronic interaction model QGSJet-II. The points at $r<0$ ($r>0$) show the variation a $\psi$-segment around $\psi = 0^\circ$ and $\psi = 180^\circ$ ($\psi = -90^\circ$ and $\psi = 90^\circ$) in the shower front plane coordinate system. The error bars are dominated by shower-to-shower fluctuations in the simulations. To obtain each profile, 10 proton showers from the library described in \secref{sec_mc_prod_full} were first normalized to one at $r=\unit{1000}{m}$ and then averaged. The geomagnetic field effect in the selected showers is maximal.}

\standardgraphic[width=0.7\textwidth]{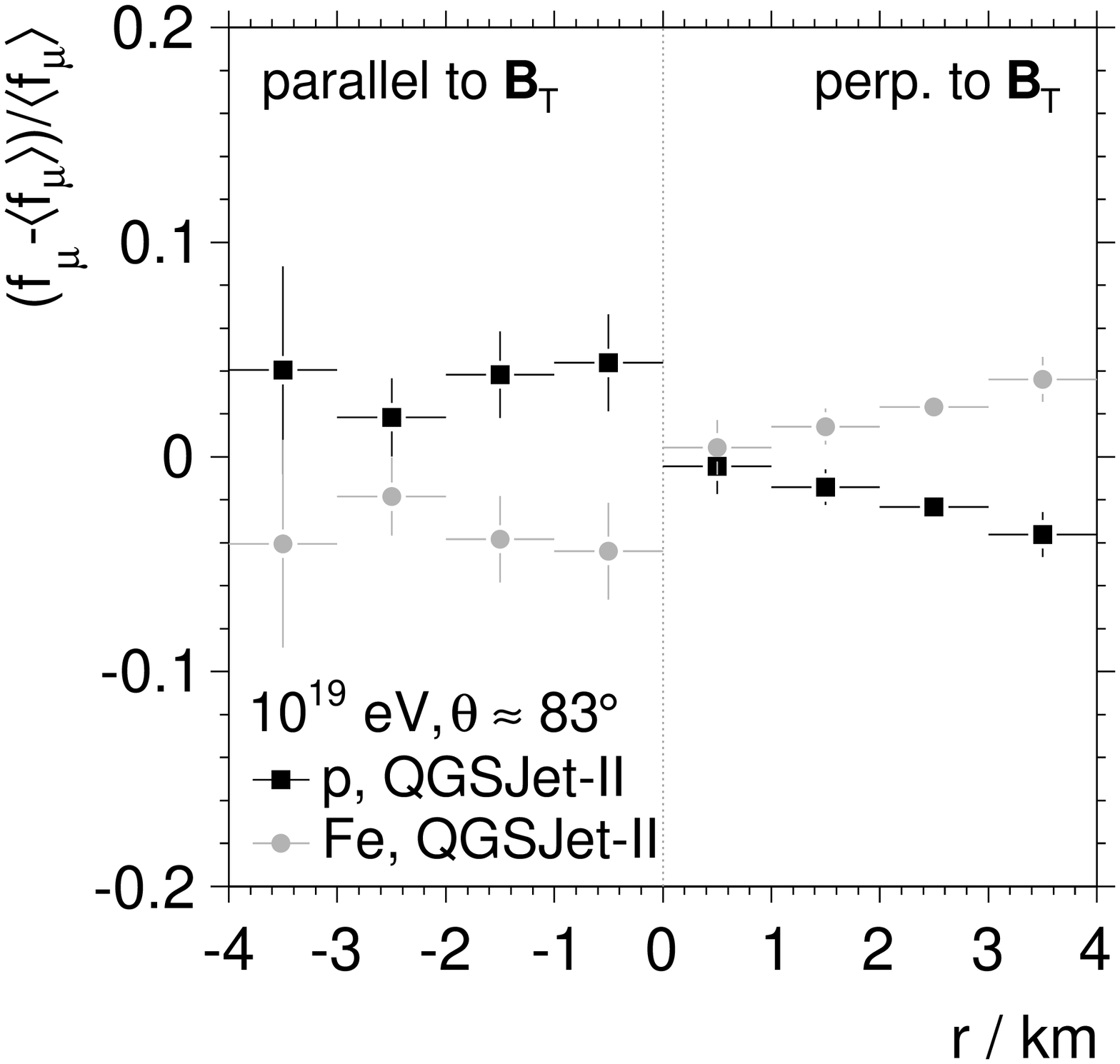}{The plot shows the same as \fg{insens_energy}, but this time the energy is kept at \unit{10^{19}}{eV} and the cosmic ray mass and the used hadronic interaction model at ultra-high energies is varied in the simulation. For each proton (iron) profile, 10 showers (2 showers) from the library described in \secref{sec_mc_prod_full} were normalized at $r=\unit{1000}{m}$ and averaged.}

This factorization approximation was already discussed qualitatively~\cite{Ave2000a,Ave2000b,Ave2002} based on simulations with the air shower code AIRES~\cite{aires} and the hadronic interaction models QGSJet01~\cite{qgsjet} and SIBYLL~\cite{sibyll}. Since it is such a central property, we confirm the approximation with the simulation code CORSIKA~\cite{corsika} using the hadronic interaction models QGSJet-II~\cite{qgsjet2} and FLUKA~\cite{fluka}.

\fg{insens_energy} and \fg{insens_mass} show that the normalized profile $f_\mu$ varies within \unit{5}{\%} between \unit{10^{18}}{eV} and \unit{10^{20}}{eV}. A variation of the cosmic ray mass $A$ between the two extreme cases of proton and iron nuclei also yields a variation of $f_\mu$ of \unit{5}{\%}.

The observed variation of $f_\mu$ can be compared to the shower-to-shower fluctuations of the total muon number at ground $N_\mu$ for cosmic protons which are at the level of \unit{13}{\%} (\unit{3}{\%}) in simulations of proton (iron) air showers with QGSJet-II. The intrinsic shower-to-shower fluctuations of $N_\mu$ are a natural limit to the obtainable energy resolution from a surface array of muon counters. The systematic impact of the factorization approximation is negligible in comparison if the cosmic ray composition is mostly light.

The theoretical uncertainties in the normalized profile $f_\mu$ are estimated by comparing~\cite{Dembinski2009} showers simulated with the hadronic interaction models QGSJet-II and EPOS~\cite{epos}. They are also at the level of \unit{5}{\%}.

The total number of muons $N'_\mu$ produced around the shower maximum depends strongly on the energy $E$ and mass $A$ of the cosmic ray. To a lesser degree, $N'_\mu$ also depends on the zenith angle $\theta$ since the charged pions decay slightly earlier into muons at larger inclinations, as stated earlier. Since the average muon energy increases with $\theta$, $N'_\mu$ has to decreases because of energy conservation. Again due to the slow variation of $d$ and $X_\text{atm} - X_\text{max}$ at large inclinations $60^\circ < \theta < 90^\circ$, the muon attenuation factor $a = N_\mu/N'_\mu$ depends only on $\theta$ in good approximation.

The form of the energy dependence of $N_\mu$ can be calculated with simplified Heitler-models of the hadronic cascade~\cite{Matthews2005} and turns out to be a power law. We can summarize these findings in another factorization:
\begin{equation}\label{eq_nmu_factorization}
  N_{\mu}(\theta,E,A) \simeq \tilde{a}(\theta) \times K(A) \times E^\beta \text{ for } \theta > 60^\circ,
\end{equation}
whereas $\beta \lesssim 1$ is a very weak function of the energy $E$ and mass $A$ of the cosmic ray and $K(A)$ depends only on the mass $A$. The energy dependency of $\beta$ can be neglected in the range $\unit{10^{18}}{eV} < E< \unit{10^{20}}{eV}$. The mass dependency is at the level of \unit{2}{\%} for a change from proton to iron cosmic rays~\cite{Ave2002,Dembinski2009}. The function $\tilde{a}(\theta)$ differs slightly from the true attenuation function $a(\theta)$ since it also parametrizes the decrease of $N'_\mu$ with $\theta$.

The theoretical uncertainties in the parameter $\beta$ and the function $\tilde{a}(\theta)$ can be propagated into $N_{\mu}$. They are at the level of \unit{10}{\%} in the range $60^\circ < \theta < 88^\circ$. The total theoretical uncertainty of $N_\mu$ is dominated by the uncertainty of the factor $K(A)$, which depends strongly on the details of hadronic interactions at ultra-high energies. The predictions for $K(A)$ of different hadronic interaction models differ at the level of about \unit{30}{\%} -- \unit{50}{\%} which is of the same order as the difference between proton and iron showers within a single model~\cite{Ave2002,Werner2007,Dembinski2009}.

We conclude that the absolute scale of the total number of muons $N_\mu$ on the ground is rather uncertain, but the normalized distribution $f_\mu$ of the muons in very inclined air showers is well-defined.

If an experiment is mostly interested in the reconstruction of $N_\mu$ from sampled muon densities with a model of $f_\mu$, we expect a theoretical uncertainty smaller than \unit{5}{\%} according to the discussed effects. A model of $f_\mu$ should provide at least the same level of precision in comparison with full air shower simulations.


\section{Monte-Carlo simulation of muon ground profiles}
\label{sec_mc_prod}

The input for the phenomenological model of the muon density $n_\mu$ on the ground is a library of simulated air showers.

In this section we shortly present the full Monte-Carlo simulations of very inclined air showers that we use in this article to derive the $n_\mu$-model. Further, we demonstrate a fast simulation approach that can be used to speed up the simulation of very inclined air showers and in particular generate a large amount of input for the model of $n_\mu$ in a short time.

\subsection{Theoretical uncertainties and statistical weight-sampling}

The modeling of hadronic particle interactions in extensive air showers is subject to considerable theoretical uncertainties. Most hadronic interactions in the shower are soft processes with small momentum transfer. So far, such interactions cannot be calculated within the fundamental theory of quantum chromodynamics.

Calculations of such interactions are based on effective theories and phenomenology which need to be fitted to data gathered in accelerator experiments. The center of mass energy in the collision of a \unit{10^{20}}{eV} proton with a nitrogen nucleus is about 400 TeV and far beyond the range of the available data. It therefore requires a bold extrapolation of these models. This explains the rather large theoretical uncertainties in contemporary air shower simulations which showed up mainly in the factor $K(A)$ of \eq{eq_nmu_factorization}.

A technical problem in detailed air shower simulations is the large number of secondary particles. A cosmic ray proton of \unit{10^{20}}{eV} produces about $10^{11}$ secondary particles. Most of them are photons, electrons, and positrons with low energies. To calculate such a shower in reasonable time, a statistical weight sampling method~\cite{Hillas1997} is applied during the shower simulation, commonly called \emph{thinning}. This procedure introduces additional artificial fluctuations to physical observables of the shower but conserves the mean values. The thinning phase is usually made as short as possible in order to keep the artificial fluctuations small compared to the natural fluctuations that appear during the shower development.

\subsection{Full simulation}\label{sec_mc_prod_full}

\begin{table*}
\centering
\capt{corsika_lib}{Parameter distribution of the proton shower library
generated with CORSIKA/\mbox{QGSJet-II}/FLUKA~\cite{corsika,qgsjet2,fluka}. The
showers in the library are distributed in small finite regions of the parameter
space. The random distribution within each region is given in the table. Each region
contains five showers. There are 1800 showers in total.}
\vspace{0.2cm}
\begin{tabular}{p{2cm} p{2.1cm} p{5.2cm} p{2.6cm}}
\hline
Parameter              & distribution               &
low edge of region & width of region \\
\hline
$\lg{E/\text{eV}}$     & flat                       & 18.0, 18.5, 19.0, 19.5,
20.0 & 0.1
 \\
$\theta/^\circ$ & $\sin(\theta)\cos(\theta)$ & 60, 70, 74, 78, 82, 86         & 2          \\
$\phi/^\circ$ & flat            & 0, 30,
60, 90, 120, 150, 180, 210, 240, 270, 300, 330 & 10  \\
\hline
\end{tabular}
\end{table*}

The analyses presented in this article are based on a set of 1800 proton showers which were generated with the CORSIKA~\cite{corsika} shower Monte-Carlo code. For the high and low energy interactions, the QGSJet-II~\cite{qgsjet2} and FLUKA~\cite{fluka} models were used, respectively.

We made a choice to distribute the showers in small finite bins in the three-dimensional parameter space of energy $E$, zenith angle $\theta$ and azimuth angle $\phi$. This leaves gaps where no showers were simulated while other parts get a higher statistical coverage. \tb{corsika_lib} summarizes the distribution of the showers. Technical details on the simulation options, the used thinning parameters and low energy cut-offs can be found in \appref{app_full_simulation}.

The calculation of one shower takes roughly \unit{1}{day} on a Pentium Xeon 2.8 GHz CPU.

\subsection{Fast simulation}\label{sec_mc_prod_fast}

The geomagnetic field $\vec{B}$ breaks the $\phi$-symmetry of the simulation of very inclined air showers and therefore the detailed simulation has to be run for each angle $\phi$. Furthermore, each of these simulation spends a lot of CPU time on the calculation of the electromagnetic cascade.

If only the muon component of the shower is of interest, it is not necessary to follow the bulk of the low energy electromagnetic particles. Photons and electrons with high enough energies are able to produce further pions in electromagnetic interactions with air nuclei. The process with the lowest energy threshold of about \unit{0.2}{GeV} is the production and decay of a $\Delta$-resonance. All electromagnetic particles with lower energies can be dropped from the simulation.

We can further exploit that the development of the hadronic cascade is partically independent of the geomagnetic field $\vec{B}$. The propagation distance of the hadronics in the field up to the shower maximum is only a small fraction of the propagation distance of the muons and the hadronic cascade develops mostly develops at very high energies. Therefore, the geomagnetic deflection can be neglected for hadrons.

This means that it is sufficient to simulate the development of the hadronic cascade only once for each combination of energy $E$ and zenith angle $\theta$ up to the point where the muons are produced. At this point the simulation can be interrupted. The muon positions and momenta can then be rotated to all desired values of $\phi$ and the detailed simulations of the muon propagation to the ground started from this point.

This idea was explored with the air shower simulation code AIRES~\cite{aires} and a custom muon propagation code. AIRES was modified in order to write out the position, energy, direction, and statistical weight of the muons at their respective production points, provided that their probability $P = e^{-d/(\gamma \beta c \tau)}$ to reach the ground is not negligible. High energy thresholds for electromagnetic particles make sure that the time-consuming calculation of the electromagnetic cascade is stopped early, see \appref{app_fast_simulation}.

The muons are then rotated to the desired direction of the shower. It is optionally possible at this point to reduce the weight-sampling of the muons in order to increase the statistical precision of the ground profile. Since the cylindrical symmetry of the shower is still intact at the production point, a muon with weight $w$ can be replaced by $n$ clones, with each a weight $w'=w/n$. A random rotation around the shower axis is applied to each clone. The cloning introduces some artificial correlations to the shower profile at the muon production altitude, but these are erased by random multiple scatterings when the muons reach the ground. The choice of $n$ is an uncritical trade-off between the statistical quality of the ground profiles and the calculation time. Values between 5 and 50 seem to be reasonable.

The custom muon propagation code implements the same physical processes for muons as the large air shower simulations codes CORSIKA or AIRES. We used a custom code to study the muon propagation effects more easily, but it is also possible to use either CORSIKA or AIRES to propagate a stack of input particles to the ground.

This method improves the calculation speed dramatically. The calculation of the hadronic cascade up to the point where the muons are produced takes about 40 minutes on a Pentium Xeon 2.8 GHz CPU. The transformation and propagation of the muons for each azimuth $\phi$ then takes only a few minutes on the same machine. If 30 values of $\phi$ are used for each pair ($E,\theta$), this leads to a total speed up factor over 1000 with respect to the simulation of the full showers while yielding practically identical results.

\section{Phenomenological model of muon ground profiles}\label{sec_map_generation}

We have seen in the previous section that it is possible to factorize the ground profile of the muon density $n_\mu$ into a normalized density profile $f_\mu$ and the total muon number $N_\mu$ on the ground (see \eq{eq_muon_density_factorized}), where $f_\mu$ is approximately independent of the cosmic ray energy $E$ and mass $A$. In this section, we will parametrize $N_\mu$ and $f_\mu$ separately as a function of the cosmic ray energy $E$, zenith angle $\theta$, and azimuth angle $\phi$, using a the proton shower library from \secref{sec_mc_prod_full} in order to derive an empirical model of the muon density $n_\mu$ on the ground.


The parametrization of $N_\mu$ is \eq{eq_nmu_factorization}. The parametrization of $f_\mu$ is more complex and done in two steps. In the first step, the normalized profile of the muon density $f_\mu^i$ of each individual shower is parametrized. In order to do this the normalized muon density $f_\mu$ is sampled in the shower front plane coordinate system. The empirical relation between $\lg f_\mu$ and $\sqrt{r}$ demonstrated by \fg{dep_thres} suggests an polynomial expansion of $\lg f_\mu$ in $\sqrt{r}$. The effects of early-late asymmetries and the geomagnetic deflections resemble dipole and quadrupole terms, respectively, which suggest a Fourier expansion in $\psi$. The sources of deviation described in \fg{divergence_muon} sum up to a smooth function. This allows one to cut of the expansions at low orders and still get a very good approximation of $f_\mu(r,\psi)$.

It should be emphasized that this approach can preserve all modulations and asymmetries found in the simulation, if enough input data is available. However, the coefficients do not have a direct physical interpretation: they just encode the physical information in a convenient way.

The expansion performed in the first step leads to a set of Fourier coefficients for every simulated shower, with a smooth dependence on the direction of the shower $(\theta,\phi)$. It is thus possible in a second step to approximate the evolution of each coefficient by an expansion in $\theta$ and $\phi$. This will be detailed in a further subsection.
 
While we focus on the parametrization of $n_\mu$, we also want to point out that the same scheme may be applied to other local observables. For example, if the response of the ground detectors depends on the energy of the muons or on their incidence angle, these quantities or their relevant combination can be parametrized in the same way.

\subsection{Parametrization of $N_\mu$}

\standardgraphic{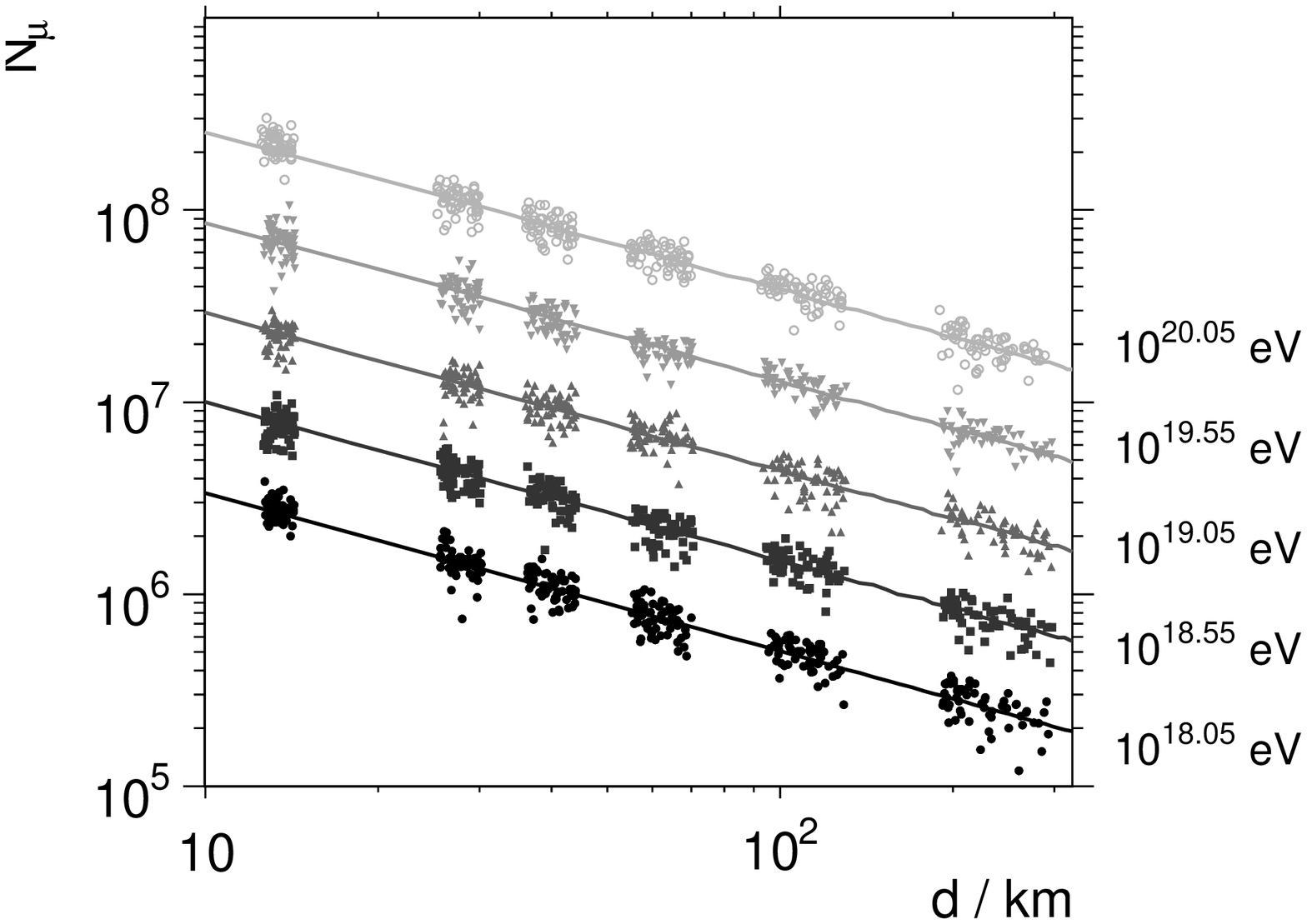}{Simulation of the total muon number on the ground $N_\mu$ as function of the distance $d$ between the shower maximum and the ground along the shower axis. Points of different colors are taken from different energy intervals, the average energy in each interval is shown at right side of the plot. The continuous lines are projections of a two-dimensional fit to the simulation (see text).}

The total number of muons on the ground $N_\mu$ can be easily extracted from each simulated shower. As nuclei give the same profile within a multiplicative factor, we regard only cosmic protons ($A = 1$), so that $N_\mu$ is only a function of zenith angle $\theta$ and the energy $E$ of the cosmic ray, as shown in \eq{eq_nmu_factorization}. A good approximation for the functional form of the energy dependency is already given in \eq{eq_nmu_factorization} and thus the only open point is the form of the attenuation function $a(\theta)$.

A good approximation to $a(\theta)$ is found empirically by drawing $N_\mu$ as a function of the distance $d(\theta)$ between the shower maximum and the
point where the shower hits the ground in log-log scale, as shown in \fg{nmu_QGSII_proton}. The relation is close to a line so that one arrives at the empirical formula:
\begin{equation}
  \lg N_\mu = D_1 + D_2 \lg(d(\theta)/\text{km}) + \beta
\lg(E/\unit{10^{18}}{eV}),
\end{equation}
where $D_1$, $D_2$, and $\beta$ are constants. For the set of proton showers simulated with QGSJet-II and FLUKA we get:
\begin{align*}
  D_1   & = 7.3043 \pm 0.0066 \\
  D_2   & = -0.8240 \pm 0.0036 \\
  \beta & = 0.9352 \pm 0.0020.
\end{align*}

\subsection{Parametrization of the normalized muon profile $f_\mu$}

The parametrization of the normalized profile of muon density $f_\mu$ is now explained in technical detail. The procedure starts with a set of simulated showers and follows four steps.
\\

\textbf{A) Coordinate transformation and density calculation.} Shower Monte-Carlos usually provide weighted muons whose positions are described in ground plane coordinates. For each shower, the ground coordinates are projected onto the shower plane coordinate system as described in \fg{coordinate_system}. To optimize the numerical precision of a polynomial expansion, it is convenient to choose a variable ranging symmetrically around zero. Furthermore, we choose $u$ in such a way that the relation between $\lg f_\mu$ and $u$ is almost a straight line. \eq{eq_nmu_vs_r} suggests:
\begin{equation}
u = 2 \frac{\sqrt{r} - \sqrt{r}_\text{min}}{\sqrt{r}_\text{max}-\sqrt{r}_\text{min}} - 1.
\end{equation}
While $r$ ranges between $r_\text{min}$ to $r_\text{max}$, $u$ ranges between $-1$ to $1$.

The shower plane is then divided into 30 bins in $\psi$ and 30 bins in $u$. The normalized density in any cell is obtained as a sum over the muons from the ground particle file falling in this cell:
\begin{equation}
f_{\mu}(\text{cell}) = \frac{\sum{w_i}}{N_{\mu}A_\text{cell}} \quad \text{with} \quad A_\text{cell} = \frac{1}{2\cos{\theta}} \: \Delta\psi\: (r_{i+1}^2 - r_i^2),
\end{equation}
where $r_i,r_{i+1}$ are the edges of the interval in $r$ corresponding to the bin in $u$ defined above, and $N_\mu$ is the total (weighted) number of muons in this shower.  

\textbf{B) Local parametrization.} The logarithmic profile $\lg f_\mu$ of the shower is now parametrized by a polynomial expansion in $u$ and a Fourier expansion in $\psi$ up to third order:
\begin{equation}
  \lg f_\mu (r,\psi) = \sum_{k=0}^{3}u^k \times \Bigl(\sum_{j=0}^{3} C_{kj} \cos(j\psi) + \sum_{j=1}^3 S_{kj} \sin(j\psi)\Bigr) \\
\end{equation}
The parametrization can be fitted to the sampled profile $\lg f_\mu$ or a shower with the linear least-squares method, reducing the fitting problem to a simple matrix inversion, and yielding coefficients which are statistically unbiased and have minimum variance. At orders higher than 3, the coefficients become statistically compatible with zero. 

\textbf{C) Global parametrization.} Each parameter $C_{kj}$ or $S_{kj}$ of the previous step is now regarded as a function of $(\theta,\phi)$ and parametrized using the whole sample of simulated showers. The dependence in $\phi$ is a Fourier expansion up to the fifth order. The dependence  in $\theta$ is a polynomial of the fifth order, again conveniently described in the range $(\theta_\text{min},\theta_\text{max})$ through the reduced variable $v = 2 (\theta - \theta_\text{min} )/(\theta_\text{max}-\theta_\text{min}) - 1$:
\begin{equation}\label{eq_param2}
\begin{split}
  C_{kj} (\theta,\phi) & = \sum_{m=0}^{5} v^m \bigg( \sum_{\ell=0}^5
  C'_{kjm\ell} \cos(\ell\phi) + \sum_{\ell=1}^5 S'_{kjm\ell}\sin(\ell\phi) \bigg) \quad \\
  S_{kj} (\theta,\phi) & = \sum_{m=0}^{5} v^m \bigg( \sum_{\ell=0}^5
  C''_{kjm\ell} \cos(\ell\phi) + \sum_{\ell=1}^5 S''_{kjm\ell}\sin(\ell\phi) \bigg) 
\end{split}
\end{equation}

The total number of parameters used to fit the normalized profile $f_\mu$ is $28 \times 66 = 1848$. This number may seem large, but only so because the model is 4-dimensional and the number of parameters per dimension multiply. The number of free parameters per dimension is small and thus the model is actually quite predictive. The parametrization is compared to one of the input showers in \fg{example0_ac}.

\multicolumngraphic{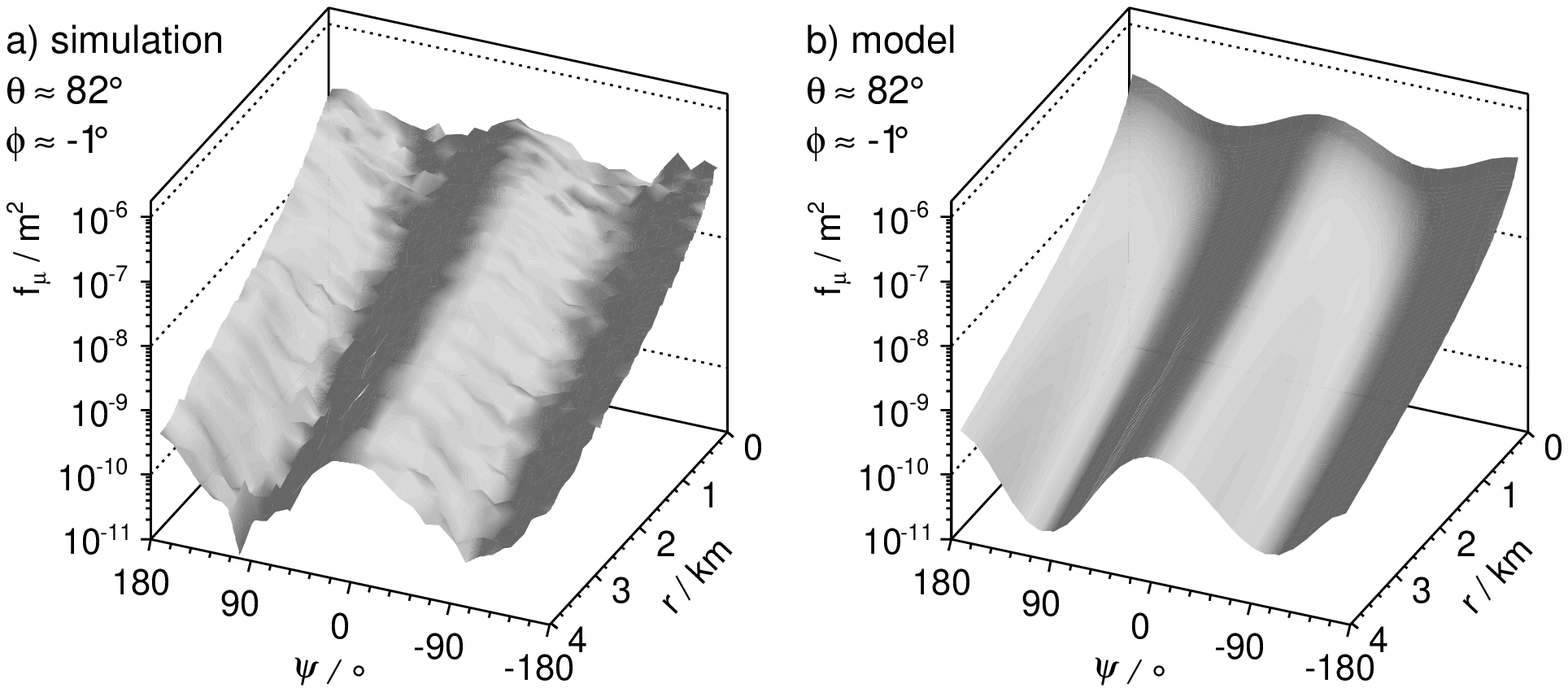}{Example of a) a simulated normalized profile $f_\mu$ of the muon density in shower front plane coordinates and b) the corresponding result produced by the parametrized model. The model reproduces the physical structures very well while ignoring the statistical fluctuations present in the simulation.}

\subsection{Precision of the parametrization}

The parametrization procedure is applied to the 1800 CORSIKA proton showers described in \tb{corsika_lib}. The bias and an estimate of the precision of the model is obtained from an analysis of the residuals of the total number of muons on the ground $(N_\mu^i - N_\mu)/N_\mu$ around the model and the distribution of the residuals of the normalized lateral profile $(f_\mu^i - f_\mu)/f_\mu$.

\multicolumngraphic{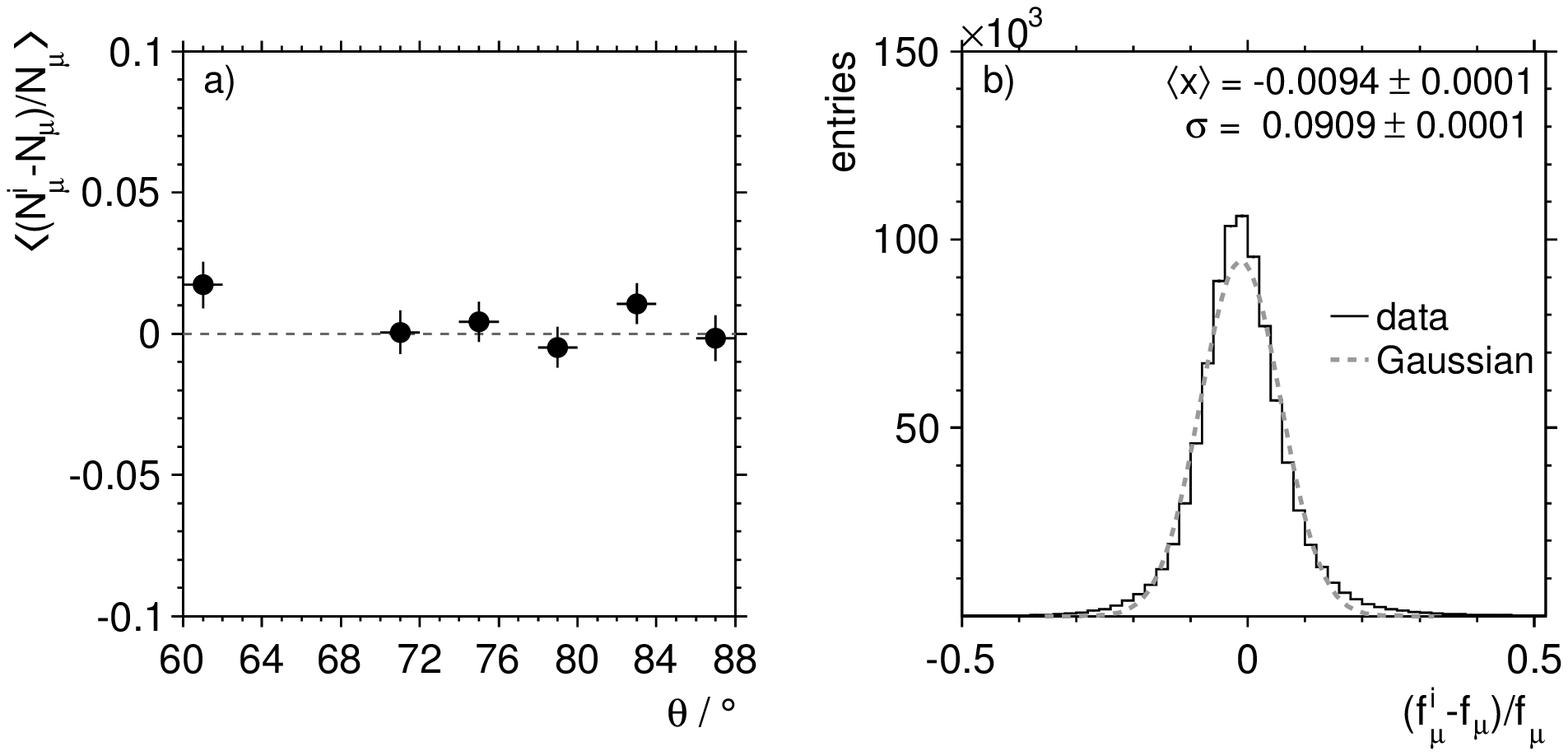}{Shown are a) the average of the residual distribution of the total muon number $N_\mu$ on the ground as a function of the zenith angle $\theta$ and b) the residual distribution (solid line) around the parametrization of the normalized muon density $f_\mu$ in the range $\unit{150}{m} < r < \unit{4000}{m}$. Also shown in b) are the mean and standard deviation of the distribution and a Gaussian fit (dashed line) for comparison.}

The analysis of the $N_\mu$-model is straight forward. The bias of the model over the whole dataset is negligible. The precision of the $N_\mu$-model is obtained by dividing the data into subsets with varying cosmic ray energy and direction and analyzing the bias $\langle (N_\mu^i - N_\mu)/ N_\mu \rangle$ of the model in these subsets. A part of the analysis is shown in \fg{map_fit_accuracy} a). A precision better than \unit{2}{\%} is observed for the $N_\mu$-model.

For the analysis of the $f_\mu$-model, the shower front plane is again divided into a grid in the coordinates $r$ and $\psi$. The grid is optimised for each individual shower in such a way that each cell contains at least 400 explicitly tracked Monte-Carlo particles with varying weights. The muons in a local cell carry roughly the same weight and thus this procedure limits the statistical fluctuations $\sigma_\text{stat}$ within in the simulated air showers to about \unit{5}{\%}. Cells which contain particles with $r<\unit{150}{m}$ are excluded from the analysis, because these particles are additionally thinned in the input simulation with a probablity $\propto r^{-1}$~\cite{corsika}. Such cells contain particle with large and strongly varying weights so that the amount of statistical fluctuation cannot be estimated in this simple fashion.

The residual distribution is shown in \fg{map_fit_accuracy}. The parametrization shows a very small bias of \unit{1}{\%} which is caused by fitting $\lg{f_\mu}$ instead of $f_\mu$ as input and the fact that $\langle \lg x \rangle \neq \lg \langle x \rangle$ for any random $x$. The data shows a relative fluctuation of \unit{9}{\%} around the model. This observed fluctuation $\sigma(f_\mu)$ is the quadratic sum of the systematic effect $\sigma_\text{model}$ caused by inaccuracies in the model and the statistical fluctuations $\sigma_\text{stat}$ in the raw data:
\begin{equation}
  \sigma^2(f_\mu) \approx \sigma_\text{model}^2(f_\mu)+\sigma_\text{stat}^2(f_\mu).
\end{equation}
Substracting $\sigma_\text{stat} \approx \unit{5}{\%}$ from the observed spread yields a precision of the $f_\mu$-model of about \unit{7}{\%}.

The simulated muon density $n_\mu = N_\mu \times f_\mu$ is therefore reproduced by the model with a global precision better than \unit{2}{\%} and a local precision better than \unit{7}{\%}. The latter is only slightly worse than the precision of the shower universality approximation from \secref{sec_energy_scaling} of about \unit{5}{\%}. To achieve such a precision for a variable with the huge dynamic range of about 8 orders of magnitude is quite remarkable.

\section{Conclusion}\label{sec_conclusion}

In this article, we have presented a practical procedure to derive a phenomenological model of the ground profile of the muon density $n_\mu$ from a large set of very inclined simulated air showers with zenith angles between $60^\circ$ and $88^\circ$. The model is based on a general parametrization of $n_\mu$ which exploits the smoothness of $n_\mu$ and the empirical observation that $n_\mu$ depends on the radial distance $r$ from the shower axis approximately like $\exp(-\alpha \sqrt{r})$. The parametrization scheme can be adapted to profiles of other ground observables.

As an example, the parametrization procedure was applied to a set of 1800 proton showers simulated with the hadronic interaction models QGSJet-II and FLUKA in the ultra-high energy range \unit{10^{18}}{eV} to \unit{10^{20}}{eV}. The derived $n_\mu$-model shows an overall bias less than \unit{2}{\%} in comparison to the simulation input. The local precision is better than \unit{7}{\%} and very close to the maximum possible accuracy of \unit{5}{\%}. The latter is the level of validity of the shower universality assumption in the regarded energy range, which is one of the foundations of the model.

We have further demonstrated a way to speed up the detailed Monte-Carlo simulation of the muon component in air showers by a large factor, keeping the whole information on the muons at ground level. The speed-up is achieved by discarding most of the calculation of the electromagnetic component and exploiting the azimuthal symmetry of the muon component around the shower axis at the production point.
 
A natural application of the $n_\mu$-model derived in this article is the reconstruction of the energy of very inclined ultra-high energy cosmic rays from data collected by a ground array of particle detectors, like the Pierre Auger Observatory. The high precision and the flexibility of the model reduce the systematic uncertainty of this reconstruction.

\section*{Acknowledgements}
This work was partly funded by the German Federal Ministry for Education and
Research BMBF and the German Research Foundation DFG. The computation of the air
shower library was supported by a grant of computing time from the
Ohio Supercomputer Center. We thank Jim Beatty and the OSC team for making the
use of this grant possible.

A part of the air shower simulations and their mass storage were possible
through a grand of computing time and storage capability at the Computing
Center in Lyon, France, to the Pierre Auger Collaboration. We like to thank the
Lyon User Support Team for their support.

For additional support and valuable discussions during the work that led to this
article, we like to thank our colleagues of the Pierre Auger Collaboration.


\appendix

\section{Full simulation: Thinning and energy thresholds}\label{app_full_simulation}

For the correct simulation of inclined showers, CORSIKA~\cite{corsika} is
compiled with the CURVED and UPWARD options. The SLANT option is not mandatory
for the analyses presented in this article, but necessary if also the longitudinal
particle profiles of the shower are to be used.

The showers are thinned according to the strategy described in ref.~\cite{Kobal2001}. The energy $E_\text{thin}$ where the thinning starts and the allowed maximum weight $w_\text{max}$ for a single Monte-Carlo particle are proportional to the primary energy $E$ of the cosmic ray.
\begin{align*}
  E_\text{thin}                  & = 10^{-6} E \\
  w_\text{max}^{\text{had},\mu}  & = 10^{-6} E/\text{GeV} \\
  w_\text{max}^{\text{e},\gamma} & = 10^{-4} E/\text{GeV} \; .
\end{align*}
The energy dependent $w_\text{max}(E)$ makes sure, that the amount of actually calculated particles is roughly independent of the primary energy $E$, which is then also true for the computation time of the showers.

As a quality/computation time trade-off, a lower maximum weight for hadrons and muons
$w_\text{max}^{\text{had},\mu}$ is chosen than for electrons and photons
$w_\text{max}^{\text{e},\gamma}$, since muons dominate in inclined showers.

The simulation drops particles below certain momentum thresholds. These thresholds have to be adapted to the detection thresholds of the regarded detectors. The following cut-offs are used
\begin{align*}
  p_\text{thres}^{\gamma}         & = 250 \text{ keV/c} \\
  p_\text{thres}^{e}              & = 250 \text{ keV/c} \\
  p_\text{thres}^\mu              & = 0.1 \text{ GeV/c} \\
  p_\text{thres}^\text{hadron}    & = 0.1 \text{ GeV/c}
\end{align*}
for photons, electrons, muons, and hadrons, respectively.

\section{Fast Simulation: Thinning and energy thresholds}\label{app_fast_simulation}

To generate the muon distribution at their production point for the fast
simulation explained in Section \ref{sec_mc_prod_fast}, the following
thinning and momentum thresholds are used in the air shower code AIRES~\cite{aires}:
\begin{equation}
  \begin{split}
    E_\text{thin}                    & = 5 \times 10^{-7} E \\
    \text{weight factor}             & = 1 \\
    E_\text{thres}^{\gamma}          & = 0.2 \text{ GeV} \\
    E_\text{thres}^{e}               & = 0.4 \text{ GeV} \\
    E_\text{thres}^{\mu}             & = 0.1 \text{ GeV} \\
    E_\text{thres}^{\text{meson}}    & = 0.15 \text{ GeV} \\
    E_\text{thres}^{\text{nucleon}}  & = 0.125 \text{ GeV} \; .
  \end{split}
\end{equation}

\end{document}